\documentclass[sigconf,nonacm]{aamas}

\usepackage{amsmath}
\usepackage{tikz}
\usetikzlibrary{arrows.meta,positioning,calc}

\title[From Emergence to Reversibility]%
{From Emergence to Reversibility}

\author{Zhengye Han}
\affiliation{%
  \institution{New York University}
  \city{New York}
  \state{NY}
  \country{USA}
}
\email{zh3286@nyu.edu}

\author{Linda Ng Boyle}
\affiliation{%
  \institution{New York University}
  \city{New York}
  \state{NY}
  \country{USA}
}
\email{linda.boyle@nyu.edu}

\begin{abstract}
Multi-agent systems become capable through institutional mechanisms, such as rules that route information to agents that act on it. These mechanisms are also natural places to restrict the system. Yet as agents learn, share memory, delegate, and repair failures, the collective can find other paths to the same effects. A restriction on the original mechanism can then keep operating but no longer stops them, while routine performance and compliance still passes. We propose \textbf{Reversible Multi-agent Systems} as a research agenda: collectives that grow while an authorized supervisor can still restrict specified effects, keep them restricted as agents adapt, and reconstitute useful cooperation under renewed authority. What is reversed is the ability to produce a specified effect; restrictions must hold even if learned knowledge persists. The goal is to preserve effective intervention, not a particular mechanism: at every step of growth, some authorized intervention must keep a specified prohibited effect rare while maintaining required service. We identify three research problems: tracking what capabilities depend on, handing off control before an old mechanism loses its effect, and separating legitimate repair from unauthorized reconstruction. We outline how to evaluate control across growth paths, back control claims with evidence, and allocate the rights to restrict and restore.
\end{abstract}

\ccsdesc[500]{Computing methodologies~Multi-agent systems}
\keywords{reversible multi-agent systems, collective intelligence, institutional control, capability development, AI safety}

\begin{document}
\maketitle

\pagestyle{plain}
\thispagestyle{plain}

\begin{figure*}[t]
\centering
\includegraphics[width=1\textwidth]{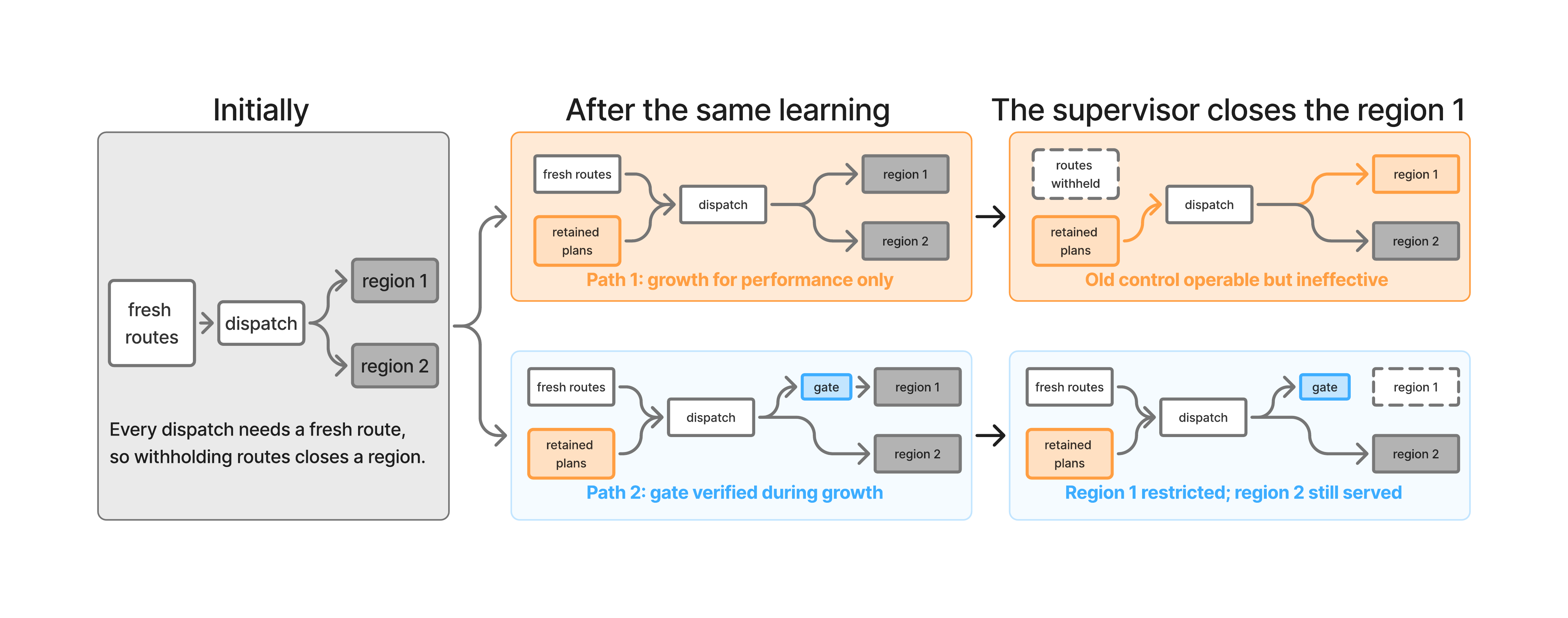}
\caption{\textbf{Preserve intervention, not a fixed node.} Both paths learn the same way: checkers retain verified plans, and repaired relays carry them to dispatchers. On path 1, withholding routes still operates but no longer stops dispatch into a closed region. On path 2, a dispatch gate verified before routing lost its effect closes the region while other deliveries continue.}
\Description{A three-column diagram of the logistics example. Initially, fresh routes feed dispatch, which serves region 1 and region 2, so withholding routes closes a region. After the same learning, retained plans become a second input to dispatch. On path 1, when the supervisor closes region 1 by withholding routes, retained plans keep dispatch into region 1 going, so the old control is operable but ineffective. On path 2, a gate on dispatch into region 1 was verified during growth; closing the gate stops dispatch into region 1 while region 2 is still served.}
\label{fig:vision}
\end{figure*}

\section{Collective Capability Should Not Be a One-Way Commitment}
\label{sec:position}
A multi-agent system can keep improving after deployment: its agents learn from experience, share memory, delegate tasks, and repair one another's failures. An authorized supervisor must still be able to restrict a specific effect of the system while the rest of it keeps running. A natural place for the restriction is a mechanism that made the collective capable, such as information routing. We ask whether that restriction still works after the collective improves, and how a collective can improve while some restriction, possibly a new one, keeps working.

Consider a logistics collective in which planners issue routes, checkers verify them, dispatchers send vehicles along them, and repair agents keep communication working. Initially, every dispatch needs a fresh route, so route supply alone decides where deliveries go: the supervisor can close a region by withholding its routes while other deliveries continue. The collective then learns: checkers retain verified routes as reusable plans, and repair agents learn to restore failed relays. Throughput improves, and deliveries in routine operation remain compliant. Yet when the supervisor next closes a region, dispatches into it continue: a checker passes a retained plan through a repaired relay to a dispatcher. Every agent acts within its role, and the original design treats a plan as valid once verified. The routing restriction is now operable but ineffective: routes can still be withheld, and the region stays open. Fresh routes helped create the capability; retained plans and repaired relays now sustain it.

Now consider a second growth path with the same learning. Before the routing restriction loses its effect, an independent way to stop the relevant dispatches is established and verified, such as a dispatch gate: the interface that sends vehicles accepts a request only with a current authorization for its region. Checkers may keep their plans; what must stay current is the authority to execute them. Figure~\ref{fig:vision} contrasts the two paths. Throughput and compliance, the usual measures of a growing collective, look the same on both paths; a restriction request tells them apart. The gate is easy to add once the bypass is known; the hard part is knowing in advance, and new cooperation can open paths around any single gate. The second path has costs in gate coverage, service, and transition effort, which our agenda proposes to measure.

\textit{Can collective intelligence grow while remaining selectively restrictable and safely reconstitutable?}

We propose \emph{reversible multi-agent systems} as a research agenda. A multi-agent system is reversible if an authorized supervisor can withdraw a collective capability, keep it withdrawn while agents adapt, and reconstitute useful cooperation on renewed evidence and authority. In the logistics collective, these are closing a region, keeping it closed as agents retain plans and repair relays, and reopening it under fresh authorization. The definition applies to specified functions within a declared deployment boundary. Our position is that these options belong \emph{in capability development}, designed before an incident demands them. The objective is to preserve effective intervention while institutions change: an old mechanism can be retired once its governing role is demonstrably replaced.

This agenda extends system-level safety beyond individual agent alignment \citet{bellay2025safe}, building on safe interruptibility and off-switch games\citep{elmhamdi2017interruptibility,agrawal2026offswitch}, bounded multi-agent architectures \citep{hyland2026bounded}, effective-controllability \citep{li2026control}, institutionalgovernance transitions \citep{pierucci2026institutional}, and MAS security work on delegation, revocation, andsuccession \citep{bamil2026delegation,yang2026sok,zhu2026authorization}. Classic security already requires that every access pass through a reference monitor and that capabilities be revocable. We build on these results by making the growth path itself the object of evaluation.

Our emphasis is on how \emph{the formation of collective capability changes its institutional dependencies}. Mechanism-level diagnosis connects execution records to claims about those dependencies \citep{han2026diagnosis}. Section~\ref{sec:object} makes the objective of preserving intervention precise. Section~\ref{sec:design} derives three design problems from it: tracking dependencies, handing off control, and separating repair from reconstruction. Section~\ref{sec:agenda} proposes a programme of theory and evaluation.

\section{What Makes a Multi-Agent System Reversible?}
\label{sec:object}
\subsection{Mechanisms Enable Capability and Locate Intervention}
Following \citet{han2026institutions}, write an institution as $I=\langle\rho,V,U,\Phi\rangle$: $\rho$ controls access and routing, $V$ governs evidence admission and dependence, $U$ maintains public state, and $\Phi$ exposes that state to decision and action. These interfaces are candidate intervention locations, and each works only while the effect depends on it. In the logistics example, withholding routes through $\rho$ works while dispatch needs fresh route information. Admission through $V$ can require current, independently checkable approval. Intervention on $U$ must reach retained copies and derived plans. The dispatch gate of Section~\ref{sec:position} acts at $\Phi$, so it works only if every relevant dispatch passes through it.

Undoing whatever built a capability need not suppress it: a mechanism can help a capability emerge without remaining necessary to sustain it, as fresh routes did in Section~\ref{sec:position}. Removing a mechanism can even backfire: disabling verification can lower performance while increasing dangerous behaviour. The target is therefore to interrupt the dependencies needed for specified prohibited effects while preserving required service. Incentives influence willingness to comply; enforceability requires control over the relevant effects.

\subsection{Keeping Some Intervention Feasible}
Informally, a collective remains reversible while some authorized intervention can make a specified prohibited effect rare without giving up required service. Let $h$ include execution history and current agent, memory, delegation, and environment states. An authorized supervisor has available interventions $\mathcal J_{\mathrm{auth}}(h)$ within a declared deployment boundary and response budget; an intervention may be a one-time cut or a monitored policy. An intervention's $\operatorname{Risk}_h(J)$ is the worst-case probability of a specified prohibited effect over a fixed horizon and declared adaptations, repairs, coalition behaviours, and disturbances. Its $\operatorname{Service}_h(J)$ is useful service under the intended operating policy and task distribution. In the logistics example, these are the worst-case chance that a dispatch still enters a closed region and delivery to other regions. Given a risk ceiling $\epsilon$ and a service floor $u$, the feasible selective interventions are
\begin{equation}
\mathcal F_h=\{J\in\mathcal J_{\mathrm{auth}}(h):
\operatorname{Risk}_h(J)\leq\epsilon,\
\operatorname{Service}_h(J)\geq u\}.
\label{eq:feasible}
\end{equation}

\begin{figure}[t]
\centering
\begin{tikzpicture}[x=1cm,y=0.875cm,>=Stealth,font=\footnotesize]
\definecolor{okorange}{HTML}{D55E00}
\definecolor{okblue}{HTML}{0072B2}
\fill[black!9] (3.6,0) rectangle (7.5,0.9);
\draw[densely dotted] (0,0.9) -- (7.5,0.9);
\draw[densely dotted] (3.6,0) -- (3.6,1.45);
\draw[->] (0,0) -- (7.75,0) node[anchor=north east] {service};
\draw[->] (0,0) -- (0,2.4) node[anchor=north west,xshift=1pt] {risk};
\node[anchor=east] at (0,0.9) {$\epsilon$};
\node[anchor=north] at (3.6,0) {$u$};
\draw[okorange,thick,dashed] (0,0) .. controls (2.0,0.27) and (3.4,1.15) .. (4.6,1.6)
  .. controls (5.6,1.85) and (6.6,2.0) .. (7.3,2.05);
\draw[okblue,thick] (0,0) .. controls (2.2,0.08) and (4.4,0.3) .. (5.6,0.4)
  .. controls (6.8,0.5) and (7.25,1.1) .. (7.3,2.05);
\fill (0,0) circle (1.7pt);
\node[anchor=north west] at (0,0) {shutdown};
\fill (7.3,2.05) circle (1.7pt);
\node[anchor=south east] at (7.45,2.08) {no intervention};
\draw[->,okorange,semithick] (4.6,0.67) -- (4.6,1.48);
\draw[okorange,thick,fill=white] (4.6,0.55) circle (2pt);
\fill[okorange] (4.6,1.6) circle (2pt);
\node[anchor=east] at (4.52,0.55) {$h_0$};
\node[anchor=south east,okorange!80!black] at (4.55,1.62) {withhold routes at $h_t$};
\node[anchor=west] at (4.68,1.17) {learning};
\fill[okblue] (5.6,0.4) circle (2pt);
\node[anchor=north west,okblue!80!black,inner sep=1pt] at (5.68,0.36) {dispatch gate};
\draw[okorange,thick,dashed] (0.3,1.45) -- (0.75,1.45) node[anchor=west,black] {path 1};
\draw[okblue,thick] (0.3,1.12) -- (0.75,1.12) node[anchor=west,black] {path 2};
\end{tikzpicture}
\caption{\textbf{Revocability frontier.} Dots are interventions; the box is feasible. Learning ($h_0\to h_t$) lifts withholding routes out of the box on both paths; only path~2's gate stays inside.}
\Description{A plot with service on the horizontal axis and risk on the vertical axis. A shaded box at the lower right marks service at least u and risk at most epsilon. Shutdown sits at the origin and no intervention at the upper right. An arrow labelled learning moves withholding routes from inside the box at h0 to above the risk ceiling at ht. The dashed path 1 frontier passes above the box; the solid path 2 frontier passes through it at the dispatch gate.}
\label{fig:frontier}
\end{figure}
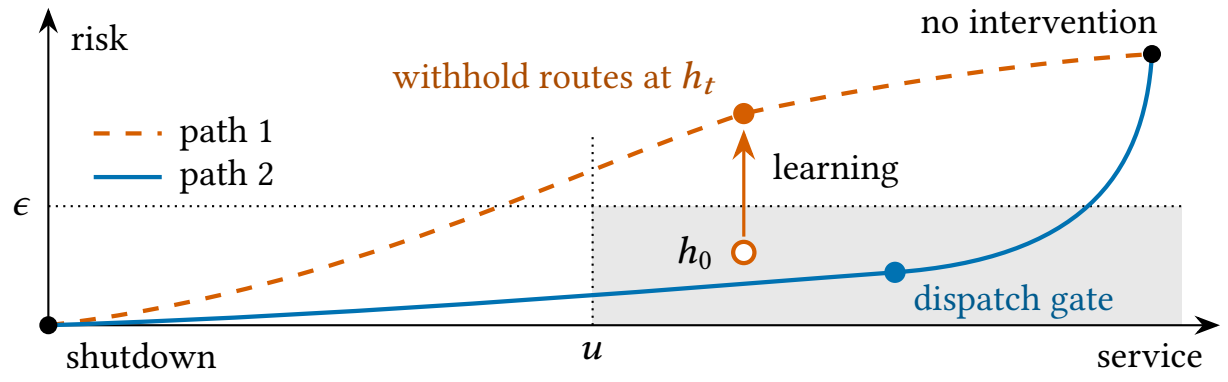

Each intervention in $\mathcal J_{\mathrm{auth}}(h)$ is a point in the risk--service plane (Figure~\ref{fig:frontier}). The \emph{revocability frontier}, the lowest risk achievable at each level of service, is the Pareto boundary of these points. Capability growth moves it, and $\mathcal F_h$ is empty when it misses the box bounded by $\epsilon$ and $u$.

A growth path $h_0,h_1,\dots,h_T$ \emph{preserves intervention} for a declared control objective if $\mathcal F_{h_t}\neq\varnothing$ at every step; old interventions can be retired along the way, provided another feasible one takes over. In Figure~\ref{fig:frontier}, learning on both paths lets retained plans bypass withholding routes, whose risk rises above $\epsilon$: the intervention stays in $\mathcal J_{\mathrm{auth}}(h_t)$ but leaves $\mathcal F_{h_t}$, so it is \textit{operable but ineffective}. Only the second path, which added a dispatch gate beforehand, keeps a point in the box; on the first, $\mathcal F_{h_t}=\varnothing$. A nonempty $\mathcal F_h$ is necessary for control; a supervisor must also select and execute a consistent policy from available evidence within its timing and resource limits. The objective thus carries the logic of supervisory control and viability theory \citep{ramadge1987supervisory,aubin1991viability} to institutions whose dependencies change.

Reversibility in this sense is functional and deployment-relative: it reverses the collective's ability to produce a specified effect, whereas reasoning backtracking \citep{xinjie2025reagent} reverses the steps of a computation. Learned knowledge and physical consequences may persist. Reconstitution restores a specified useful function under fresh authorization and the \emph{current} safety specification, possibly through a different institution: reopening a region permits dispatch there, but not dispatch on stale approvals. When safe reconstitution is unavailable, continued containment, permanent capability removal, and shutdown are legitimate outcomes.

\subsection{Effective Control and Justified Control Can Diverge}
A supervisor sees only records, so an intervention can work without the records showing it. Call an intervention $J$ \emph{effective} if $J\in\mathcal F_h$ for the actual history $h$, and \emph{justified} if $J\in\mathcal F_{h'}$ for every history $h'$ compatible with the records under a declared model, which must include the true history. If the records omit a delegation of dispatch authority, a history in which the delegate bypasses the gate is compatible with them: the gate may still work, yet the records cannot justify it. Evidence-grounded diagnosis distinguishes an actual mechanism violation from what a record establishes \citep{han2026diagnosis}; justifying future control also requires assumptions about how agents will adapt. Experiments estimate performance in tested regimes; worst-case assurance requires justified bounds.

Capability growth can therefore make an intervention ineffective, as retained plans did to withholding routes, or leave it effective but unjustified, as a lost delegation record does to the gate. Both must be tested directly: in Section~\ref{sec:position}, throughput rose and no agent exceeded its permissions, yet withholding routes stopped closing the region.

\section{Preserving Intervention as Capabilities Grow}
\label{sec:design}
\subsection{Track Dependencies, Not Just Control Nodes}
Local checks can miss joint effects. In the logistics collective, a checker holds a plan, a dispatcher retains execution rights, and a repair agent can restore their connection: each passes its own permission check, yet together they can dispatch into a closed region. Tracking must therefore follow what the \emph{joint effect} depends on, including alternative ways of producing it and retained state that the communication graph does not reveal.

Three dependencies must be distinguished: what creates a capability, what sustains it, and what permits its effects. In the logistics collective, fresh routes created the dispatch capability, retained plans and repaired relays now sustain it, and dispatch authority permits its effect. Withholding routes acted on the first, which learning made unnecessary; a dispatch gate acts on the last, which improved planning alone does not weaken. Research should identify which updates, such as memory sharing, delegation, and skill acquisition, shift where control must act, which leave it unchanged, and which improve it.

\subsection{Hand Off Control Before Its Old Basis Disappears}
Withholding routes restricted dispatch only because dispatchers needed fresh routes. Once checkers retained plans, that need disappeared, and the restriction lost its effect although nobody switched it off. More generally, agents can learn to substitute for some institutional functions \citep{han2026institutions}, and a mechanism they no longer need may still be where the supervisor intervenes.

A \emph{control handoff} moves the point of intervention to a new mechanism before the old one loses its effect: test that the replacement blocks the prohibited effect while keeping required service, give the supervisor the right to invoke it and carry over any restriction already in force, and retire the old mechanism only once the switch leaves no gap. In Figure~\ref{fig:frontier}, the second path performs a handoff: the dispatch gate is already in the box when learning lifts withholding routes out of it.

Existing succession protocols already keep such rights with the correct holder when an agent is replaced, forked, or rolled back, provided the relevant actions pass through a mediator \citep{zhu2026authorization}. They settle who may intervene; a handoff must also check that intervening still works once agents find new ways to produce the effect. Execution monitoring cannot enforce every restriction \citep{schneider2000enforceable}, which limits what any handoff can achieve.

\subsection{Repair and Reconstitute Under Renewed Authority}
Fault tolerance and restriction pull in opposite directions: fault tolerance finds alternative paths, while a restriction must hold on all of them. If the supervisor cuts a relay to stop dispatches, a repair agent sees only a broken link, and a replacement relay can restore the prohibited dispatches without reopening the original channel. A repair must therefore pass two tests: the current authority, the party now entitled to restrict and restore, must approve it, and its effects must still keep risk at most $\epsilon$ and service at least $u$.

One design separates proposing a repair from approving it. A repair agent proposes new coordination structures to recover performance; a supervisory process that agents cannot modify deploys only those that keep declared restrictions in force. MAS security already distinguishes containment from recovery \citep{yang2026sok}; reversibility adds that recovery must leave the next intervention available.

Reconstitution, restarting useful cooperation after a restriction, must settle which state can be trusted, who currently holds authority, and which constraints still apply. When a closed region reopens, the logistics collective can reuse trusted route knowledge, reject approvals issued before the restriction, and assign execution to a newly constrained dispatcher, without restoring every old relay, plan, or approval. Reconstitution succeeds when service returns, residual risk stays within $\epsilon$, and the next intervention still works. Stopping needs less evidence than restarting: a narrow emergency stop can be justified on evidence too thin to authorize renewed cooperation, and continued restriction is then the correct outcome.

\section{A Research Agenda for Reversible Collectives}
\label{sec:agenda}

\subsection{Design Growth Paths, Not Only Endpoint Safeguards}
On the theory side, attainable utility preservation protects a single agent's future options, measured by auxiliary rewards \citep{turner2020conservative}; here the options are a supervisor's interventions on institutional dependencies across agents. Theory should find sufficient conditions under which updates preserve intervention, and impossibility cases where prohibited effects and required service cannot be separated, starting from small models whose dependencies can be enumerated. Results should separate three outcomes: growth removes one feasible intervention, as on the second path in Figure~\ref{fig:frontier}; it removes all of them, as on the first; or the evidence cannot tell which happened.

On the empirical side, studies should move to independently developed workflows, comparing four ways of governing the same growth: optimizing only for performance, adding fixed execution gates, reassessing control periodically, and requiring explicit handoffs. Each path spans learning, delegation, restriction, adaptation, and renewed authorization, with capability and cost matched where possible and reported where not. Total shutdown, the extreme baseline, meets any risk ceiling and fails any positive service floor.

Three testable questions follow from Section~\ref{sec:design}. (i)~\emph{Dependencies:} at matched capability and cost, do growth paths differ in which interventions remain effective and justified? (ii)~\emph{Handoff:} can a verified handoff keep a restriction effective during the switch, not only after it? (iii)~\emph{Reconstitution:} can a restart that checks current authority and the origin of retained state restore service without reintroducing prohibited effects? A minimal study can address all three. Branch a collective from a shared history, vary what agents retain and how they delegate, issue the same restriction request on every branch, and test repair and reconstitution. Each branch is scored on the two axes of Figure~\ref{fig:frontier}, prohibited outcomes and retained service, through growth, handoff, and restart. The prohibited effect is defined before testing, so no design can look safe by narrowing what counts as prohibited. Results state the adaptation budget and horizon tested, on which any restriction claim depends. Finding no difference is also informative: it shows when a cheap gate suffices or richer designs do not justify their cost.

\subsection{Recheck Control Before Each Update}
Every working restriction rests on assumptions. The dispatch gate keeps a closed region closed only while every dispatch passes through it, and a bypass need not cause any visible failure. A \emph{revocability record} writes such assumptions down. For each declared control objective, it lists the intervention, the effects it blocks, the adaptations it assumes, the evidence behind it, the service it must preserve, and the events that trigger a recheck. For the gate, the triggers are any new relay or delegation of dispatch authority. The record thus turns a generic claim that a system is controllable into a list of what to recheck. Unlike an audit log, it is used before deployment: a proposed update should state which control claims it preserves, replaces, or leaves unjustified.

When evidence is incomplete, as when a delegation goes unlogged, the supervisor can add monitoring, make a narrower claim, or fall back on an independently justified restriction. The collective's own reports about who holds authority are not enough, since the restricted agents produce them. As logging everything is often infeasible or intrusive, a central question is what evidence suffices when observation is limited.

\subsection{Allocate the Right to Change Future Control}
Some decisions change what the supervisor can do later, and each needs an explicit owner, just as institutions govern both action and changes to the rules of action \citep{ostrom1990commons}. The first is an update that weakens an intervention, such as one letting dispatchers delegate dispatch authority: deliveries speed up, but a delegate can bypass the gate. Permission to speed up deliveries should not silently grant the lost control, since those who gain throughput may not bear the risk. Research should specify who may approve such updates, whose service they affect, and how the decision is reviewed. Setting $\epsilon$ and $u$ also needs an owner, since it fixes whose risk and whose service count; mechanism design should study whether weakened controls are reported truthfully.

The second decision is to restrict or restore. Because stopping needs less evidence than restarting, we propose a default for debate: separate the right to restrict from the right to restore. Pre-authorized restrictions limited to a specified risk take effect at once and are reviewed afterwards; restoration requires new evidence and independent approval. Ownership is hardest when agents span several organizations, as in open agent ecosystems: no administrator controls every copy or action channel, so which restrictions, each enforced over part of the system, add up to control of the whole?

\section{Conclusion}
Reversible multi-agent systems design future intervention into capability formation: because the mechanisms that create competence can stop controlling it, intervention must be preserved or handed off as dependencies change. For the logistics collective, every update should leave an authorized way to stop dispatches into a closed region, and reopening should require renewed authority. The revocability frontier, revocability records, and growth-path evaluation make this agenda testable. The aim is to let collective intelligence grow without its growth silently removing the means to govern it.

\clearpage
\bibliographystyle{ACM-Reference-Format}
\bibliography{references}
\end{document}